\documentclass{appolb}
\usepackage{graphicx}

\begin{document}

\title{ Neutrino Processes  with Hot  Nuclei in Supernovae%
\thanks{Presented at the Zakopane Conference on Nuclear Physics ``Extremes of the Nuclear
Landscape'', Zakopane, Poland, August 28  -- September 4, 2016} }
\author{Alan A. Dzhioev, A. I. Vdovin
\address{Bogoliubov Laboratory of Theoretical Physics, JINR, Dubna, Russia}
}
\maketitle
\begin{abstract}
In this paper, we calculate cross sections for charged-current
neutrino-nucleus processes occuring under presupernova conditions.
To treat thermal effects we extend self-consistent Skyrme-QRPA
calculations to finite  temperature by using the formalism of thermo
field dynamics. The numerical results are presented for the sample
nuclei, $^{56}$Fe and $^{82}$Ge.
\end{abstract}

\section{Introduction}

Neutrino reactions  play a central role during the collapse and
shock revival phase  of core-collapse supernova~\cite{Janka2007}.
The reliable calculations of  neutrino-nucleus cross sections
requires a detailed knowledge of the Gamow-Teller (GT) strength in
nuclei. The problem is rather complicated, since the
finite-temperature supernova environment implicitly demands to
consider GT transitions between thermally excited nuclear states.
Such transitions remove the reaction threshold and dominate the
low-energy cross sections.

In~\cite{Sampaio}, the neutrino-nucleus cross sections were computed
using the  large-scale shell model  (LSSM) diagonalization approach.
Although the LSSM approach provides a detailed GT strength
distribution for the nuclear ground and lowest excited states, it
partially employs the Brink hypothesis when treating GT transitions
from high-lying excited states. In addition, present computer
capabilities allow LSSM calculations only for iron-group nuclei
($A\le 65$), whereas neutrino reactions with heavier mass and
neutron-rich nuclei also may play important role in core-collapse
supernovae. Here, we present the alternative method to account for
thermal effects on neutrino-nucleus reactions.

\section{Formalism}

In~\cite{Dzhioev}, thermal effects on the cross sections of
neutrino-nucleus  reactions were studied by QRPA calculations
extended to finite temperature by the thermo field dynamics
formalism (TQRPA)\cite{Umezawa,Dzhioev3}. The techniques does not
rely on the Brink hypothesis and allows to calculate the thermal
strength function  in a thermodynamically consistent way, i.e.,
without violating the principle of detailed balance.
In~\cite{Dzhioev}, the TQRPA calculations were based on a nuclear
Hamiltonian with  locally adjusted parameters.
In~Ref.~\cite{Dzhioev2}, a self-consistent approach combining the
TQRPA with the Skyrme energy density functional has been introduced
and  applied to neutral-current neutrino-nucleus reactions. In the
present note we use the Skyrme-TQRPA to predict the cross sections
for $(\nu_e,e^-)$ and $(\overline{\nu}_e,e^+)$ reactions with
$^{56}$Fe and $^{82}$Ge.

\section{Results}

\begin{figure}[t]
\centerline{%
\includegraphics[width=0.65\textwidth]{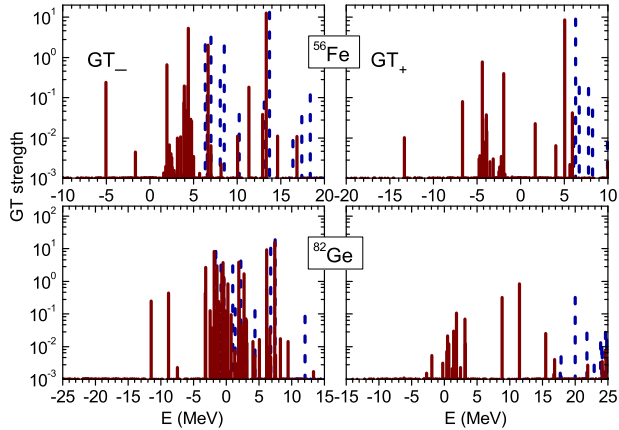}}
\caption{The GT$_-$ (left panels) and GT$_+$ (right panels) strength
functions for $^{56}$Fe and $^{82}$Ge calculated at $T=0$ (dashed
peaks) and $T=1.72$~MeV (solid peaks).
 }
\label{GT}
\end{figure}

In Fig.~\ref{GT}, we display on a logarithmic scale the GT$_-$ and
GT$_+$ distributions  in   $^{56}$Fe and $^{82}$Ge nuclei calculated
with the SLy4 forces for $T=0$ (i.e. for the ground state) and
$T=1.72~\mathrm{MeV}$ ($2\times 10^{10}~\mathrm{K}$). The
parametrization SLy4 rather well reproduces the experimental
positions of the GT$_-$ and GT$_+$ resonances in $^{56}$Fe. Because
the Brink hypothesis is not fulfilled within the TQRPA, the GT
strength functions change with temperature. Thermal effects are most
significant for the GT$_+$ distribution in $^{82}$Ge, where the
temperature rise lowers the resonance peak by about 8~MeV. As shown
in~\cite{Dzhioev}, the physical origin of this significant downward
shift is the interplay between two unblocking mechanisms of GT$_+$
transitions in neutron-rich nuclei: configuration mixing induced by
pairing correlations and thermal excitations. Thermal excitations
also unblock low- and negative-energy GT transitions. The strength
of negative-energy transitions exponentially increases with
temperature in accordance with the principle of detailed
balance~\cite{Dzhioev}. Nevertheless, the TQRPA preserves the Ikeda
sum rule.

\begin{figure}[t]
\centerline{%
\includegraphics[width=0.65\textwidth]{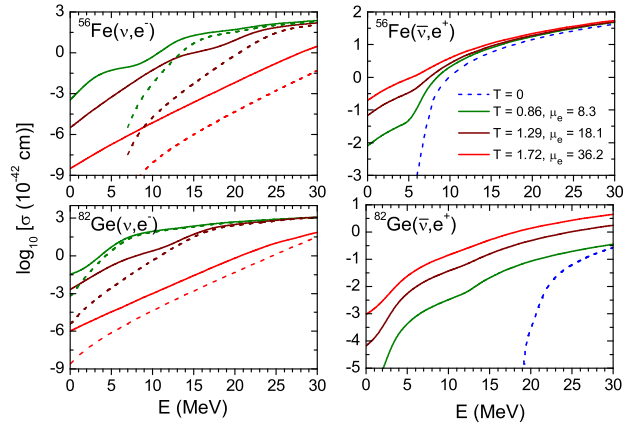}}
\caption{Cross sections of absorption reactions of $\nu_e$ and
$\bar{\nu}_e$ by $^{56}$Fe and $^{82}$Ge. The cross sections are
computed at finite temperature and finite chemical potential (both
in MeV). For comparison the ground state results are shown by the
dashed lines. } \label{CrSect}
\end{figure}

Figure~\ref{CrSect} shows the cross sections obtained with the GT
thermal strength functions computed using TQRPA with the SLy4 force.
In the collapsing core, the chemical potential of the degenerate
electron gas increases faster than the temperature. Therefore,
neutrino absorption cross sections are drastically reduced for
low-energy neutrinos due to electron blocking in the final states.
In contrast, antineutrino absorption cross sections are  strongly
enhanced at finite temperature. For $^{56}$Fe, this enhancement is
mostly caused by negative-energy GT$_+$ transitions from thermally
excited states. Such transitions dominate the cross section for
$E_{\overline{\nu}}<5~\mathrm{MeV}$. In $^{82}$Ge, negative-energy
transitions are suppressed  and the observed cross section
enhancement reflects the downward shift of the GT$_+$ resonance and
the thermal unblocking of low-energy transitions.

\begin{figure}[t]
\centerline{%
\includegraphics[width=0.65\textwidth]{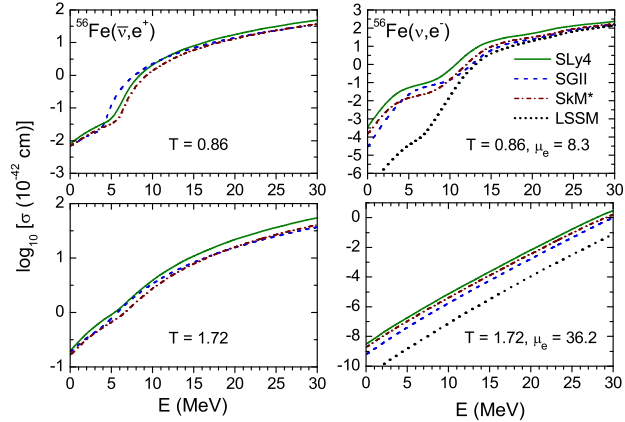}}
\caption{Finite temperature cross sections for
$^{56}\rm{Fe}(\nu_e,e^-)$ $^{56}\rm{Fe}(\bar{\nu}_e,e^+)$ reactions
computed with different Skyrme forces. For $\nu_e$ absorption the
LSSM results are also shown~\cite{Sampaio}. The temperature and chemical potential are in MeV.} \label{CrSect2}
\end{figure}

In Fig.~\ref{CrSect2}, we compare the finite temperature cross
sections for $^{56}$Fe obtained with the SLy4, SGII and SkM* Skyrme
parametrizations. As seen from the plots, the spread in the cross
sections computed with different Skyrme forces is less than an order
of magnitude. We also note the Skyrme-TQRPA results are noticeable
larger than the LSSM ones and the discrepancy reduces with
increasing neutrino energy. As shown in~\cite{Dzhioev}, the reason
for this discrepancy is that  shell-model calculations are partially
based on the Brink hypothesis when treating GT transitions from
excited states and therefore underestimate the contribution of low-
and negative-energy thermally unblocked transitions.

\section{Summary}

Cross sections for (anti)neutrino absorption by hot nuclei in the
supernova environment were calculated for  $^{56}$Fe and $^{82}$Ge
within the Skyrme-TQRPA approach. Temperature-driven changes in the
cross sections were explained by considering thermal effects on the
GT$_\pm$ strength distributions. It was found that different Skyrme
forces predict cross sections which do not differ significantly.
This could indicate a robustness of the results against the
variation of the Skyrme force parameters.

\end{document}